\documentclass[conference]{IEEEtran}

\usepackage{amsmath,amsfonts,amssymb,amsthm,cite}

\usepackage{algorithmic}
\usepackage{algorithm}
\newcommand{\INDSTATE}[1][1]{\STATE\hspace{#1\algorithmicindent}}

\def\N{\mathbb{N}}

\def\RR{\mathbb{R}}

\def\e{\varepsilon}
\def\d{\delta}

\def\1{\mathds{1}}

\def\P{\mathcal{P}}
\def\R{\mathcal{R}}
\def\Q{\mathcal{Q}}

\def\H{\mathop{H}}
\def\h{\mathop{h}}

\def\sh{\mathop{sh}}

\newtheorem{observation}{Observation}[section]
\newtheorem{theorem}{Theorem}[section]

\newtheorem{lemma}{Lemma}[section]
\newtheorem{corollary}{Corollary}[section]

\theoremstyle{definition}

\newtheorem{example}{Example}[section]
\newtheorem{problem}{Problem}[section]

\begin{document}
%
\title{Partition Reduction for Lossy Data Compression Problem}

\author{\IEEEauthorblockN{Marek \'Smieja}
\IEEEauthorblockA{Institute of Computer Science\\
Department of Mathematics and Computer Science\\
Jagiellonian University\\
Lojasiewicza 6, 30-348, Krakow, Poland\\
Email: marek.smieja@ii.uj.edu.pl}
\and
\IEEEauthorblockN{Jacek Tabor}
\IEEEauthorblockA{Institute of Computer Science\\
Department of Mathematics and Computer Science\\
Jagiellonian University\\
Lojasiewicza 6, 30-348, Krakow, Poland\\
Email: jacek.tabor@ii.uj.edu.pl}
}

\maketitle

\begin{abstract}
We consider the computational aspects of lossy data compression problem, where the compression error is determined by a cover of the data space. We propose an algorithm which reduces the number of partitions needed to find the entropy with respect to the compression error. In particular, we show that, in the case of finite cover, the entropy is attained on some partition. We give an algorithmic construction of such partition.
\end{abstract}


%
\IEEEpeerreviewmaketitle

\section{Introduction}
The basic description of lossy data compression consists of the quantization of the data space into partition and (binary) coding for this partition. Based on the approach of A. R\'enyi's \cite{Re1, Re2} and E. C. Posner et al. \cite{Po1, Po2, Po3}, we have recently presented an idea of the entropy which allows to combine these two steps \cite{Sm}. The main advantage of our description over classical ones is that we consider general probability spaces without metric. It gives us more freedom to define the error of coding.

In this paper we concentrate on the calculation of the entropy defined in \cite{Sm}. We propose an algorithm which allows to reduce drastically the computational effort to perform the lossy data coding procedure. 

To explain precisely our results let us first introduce basic definitions and give their interpretations. In this paper, if not stated otherwise, we always assume that $(X, \Sigma, \mu)$ is a subprobability space\footnote{We assume that $(X,\Sigma)$ is a measurable space and $\mu(X) \leq 1$.}. As it was mentioned, the procedure of lossy data coding consists of the quantization of data space into partition and binary coding for this partition. We say that family $\P$ is a partition if it is countable family of measurable, pairwise disjoint subsets of $X$ such that
\begin{equation}
\mu(X \setminus \bigcup_{P \in \P} P) = 0.
\end{equation}
During encoding we map every given point $x \in X$ to the unique $P \in \P$ if and only if $x \in P$. Binary coding for the partition can be simply obtained by Huffman coding of elements of $\P$.

The statistical amount of information given by optimal lossy coding of $X$ by elements of partition $\P$ is determined by the entropy of $\P$ which is \cite{Sh}:
\begin{equation}
\h(\mu; \P) := \sum_{P \in \P} \sh(\mu(P)),
\end{equation}
where $\sh(x):=-x\log_2(x)$, for $x \in (0,1]$ and $\sh(0):=0$ is the Shannon function.

The coding defined by a given partition causes specific level of error. To control the maximal error, we fix an error-control family $\Q$ which is just a measurable cover of $X$. Then we consider only such partitions $\P$ which are finner than $\Q$ i.e. we desire that for every $P \in \P$ there exists $Q \in \Q$ such that $P \subset Q$. If this is the case then we say that $\P$ is $\Q$-acceptable and we write $\P  \prec \Q$.

To understand better the definition of the error-control family let us consider the following example. 
\begin{example} \label{prz1}
Let $\Q_\e$ be a family of all intervals in $\RR$ with length $\e>0$. Every $\Q_\e$-acceptable partition consists of sets with diameter at most $\e$. Then, after encoding determined by such partition, every symbol can be decoded at least with the precision $\e$. The above error-control family was considered by A. R\'enyi \cite{Re1, Re2} in his definition of the entropy dimension. As the natural extensions he also studied the error-control families built by all balls with given radius in general metric spaces. Similar approach was also used by E. C. Posner \cite{Po1, Po2, Po3} in his definition of $\e$-entropy\footnote{E. C. Posner considered in fact $(\e,\d)$-entropy which differs slightly from our approach.}.

In the case of general measures, it seems to be more natural to vary the lengths of intervals from the error-control family. Less probable events should be coded with lower precision (longer intervals) while more probable ones with higher accuracy (shorter intervals). Our approach allows to deal easily with such situations. 
\end{example}

To describe the best lossy coding determined by $\Q$-acceptable partitions, we define the entropy of $\Q$ as
\begin{equation}
\begin{array}{l}
\H(\mu;\Q) := \\[0.4ex]
\inf\{\h(\mu;\P) \in [0;\infty] : \P \text{ is a partition and } \P \prec \Q\}.
\end{array}
\end{equation}

Let us observe what is the main difficulty in the application of this approach to the lossy data coding:
\begin{example} \label{exCom}
Let us consider the data space $\RR$ and the error-control family $\Q=\{(-\infty,1],[0,+\infty)\}$. In such simple situation there exists uncountable number of $\Q$-acceptable partitions which have to be considered to find $\H(\mu;\Q)$.
\end{example}

In this paper, we show how to reduce the aforementioned problem to the at most countable one. In the next section, we propose an algorithm which for a given partition $\P \prec \Q$, allows to construct $\Q$-acceptable partition $\R \subset \Sigma_\Q$ with not greater entropy than $\P$, where $\Sigma_\Q$ denotes the sigma algebra generated by $\Q$ (see Algorithm II.1).

As a consequence we obtain that the entropy $\H(\mu;\Q)$ can be realized only by partitions $\P \subset \Sigma_\Q$ (see Corollary \ref{def}). In the case of finite error-control families $\Q$, we get an algorithmic construction of optimal $\Q$-acceptable partition. More precisely, if $\Q$ is an error-control family then there exists $k$ sets $Q_1,\ldots,Q_k \in \Q$ such that (see Corollary \ref{corAttained}):
\begin{equation}
\H(\mu;\Q) = \h(\mu;\{Q_i \setminus \bigcup_{j=1}^i Q_j \}_{i=1}^k ).
\end{equation}

\section{Algorithm for Partition Reduction}

In this section we present an algorithm which for a given $\Q$-acceptable partition $\P$ constructs $\Q$-acceptable partition $\R \subset \Sigma_\Q$ with not greater entropy. We give the detailed explanation that $\h(\mu;\R) \leq \h(\mu;\P)$.

We first establish the notation: for a given family $\Q$ of subsets of $X$ and set $A \subset X$, we denote:
\begin{equation}
\Q_A = \{ Q \cap A: Q \in \Q \}.
\end{equation}

Let $\Q$ be an error-control family on $X$ and let $\P$ be a $\Q$-acceptable partition of $X$. We build a family $\R$ according to the following algorithm: 
\smallskip \hrule \smallskip
\textbf{Algorithm II.1.}\smallskip \hrule \smallskip
\begin{flushleft}
\begin{algorithmic}                    
\STATE {\bf initialization}
\INDSTATE $i:=0$
\INDSTATE $X_0:=X$
\INDSTATE $\R:=\emptyset$
\WHILE{$\mu(X_i) > 0$} 
	\STATE {\textit{Let} $P^i_{\max} \in \P_{X_i}$ \textit{be such that}} 
   \INDSTATE {$\mu(P^i_{\max}) = \max\{ \mu(P): P \in \P_{X_i} \}$}
   \STATE {\textit{Let} $R_i \in \Q_{X_i}$ \textit{be an arbitrary set}}
   \INDSTATE{\textit{which satisfies} $P^i_{\max} \subset R_i$}
   \STATE{$\R=\R \cup \{R_i\}$}
   \STATE{$X_{i+1}:=X \setminus (R_1 \cup \ldots \cup R_i)$}
   \STATE{$i:=i+1$}
\ENDWHILE
\end{algorithmic}
\end{flushleft}\smallskip    
\hrule
\smallskip

We are going to show that Algorithm II.1 produces the partition $\R$ with not greater entropy than $\P$. Before that, for the convenience of the reader, we first recall two important facts, which we will refer to in further considerations. 
\begin{observation} \label{shProperty}
Given numbers $p \geq q \geq 0$ and $r > 0$ such that $p, q, p+r, q-r \in [0,1]$, we have:
\begin{equation}
	\sh(p) + \sh(q) \geq \sh(p+r) + \sh(q-r).
\end{equation}
\end{observation}
\begin{IEEEproof}
For the proof we refer the reader to \cite[Section 6]{Sh} where similar problem is illustrated for $p+q=1$.
\end{IEEEproof}

\begin{flushleft}
\textbf{Consequence of Lebesgue Theorem} (see \cite{Ki}) \textit{Let $g:\N \rightarrow \RR$ be summable i.e. $\sum\limits_{k \in \N} g(k) < \infty$ and $\{f_n\}_{n = 1}^\infty$ be a sequence of functions $\N \rightarrow \RR$ such that $|f_n| \leq g$, for $n \in \N$. If $f_n$ is pointwise convergent, for every $n \in \N$, then $\lim\limits_{n \to \infty} f_n$ is summable and}
\begin{equation}
	\sum_{k \in \N} \lim_{n \to \infty} f_n (k)= \lim_{n \to \infty} \sum_{k \in \N} f_n(k).
\end{equation}
\medskip
\end{flushleft} 

Let us move to the analysis of Algorithm II.1. We first check what happens in the single iteration of the algorithm.
\begin{lemma} \label{lemma}
We consider an error-control family $\Q$ and a $\Q$-acceptable partition $\P$ of $X$. Let $P_{\max} \in \P$ be such that:
\begin{equation}
\mu(P_{\max}) = \max\{ \mu(P): P \in \P \}.
\end{equation}
If $Q \in \Q$ is chosen so that $P_{\max} \subset Q$ then
\begin{equation} \label{nierLem}
\h(\mu;\{Q\} \cup \P_{X \setminus Q} ) \leq \h(\mu;\P).
\end{equation}
\end{lemma}
\begin{IEEEproof}
Clearly, if $\h(\mu;\P) = \infty$ then the inequality (\ref{nierLem}) holds trivially. Thus we assume that $\h(\mu;\P) < \infty$. 

Let us observe that it is enough to consider only elements of $\P$ with non zero measure -- the number of such sets can be at most countable. Thus, let us assume that $\P = \{P_i\}_{i=1}^\infty$ (the case when $\P$ is finite can be treated in similar manner). 

For simplicity we put $P_1:=P_{\max}$. For every $k \in \N$, we consider the sequence of sets, defined by
\begin{equation}
Q_k:= \bigcup_{i=1}^k (P_i \cap Q ).
\end{equation}
Clearly, for $k \in \N$, we have
\begin{equation}
Q_1 = P_1,
\end{equation} 
\begin{equation}
Q_k \subset Q_{k+1},
\end{equation}
\begin{equation} \label{zawier1}
P_i \cap Q_k  = P_i \cap Q \text{, for } i \leq k,
\end{equation}
\begin{equation} \label{zawier2}
P_i \cap Q_k = \emptyset \text{, for } i > k,
\end{equation}
\begin{equation} \label{zawier3}
\lim_{n \to \infty} \mu(Q_n) = \mu(Q).
\end{equation}

To complete the proof it is sufficient to derive that for every $k \in \N$, we have:
\begin{equation}\label{pierwsza}
\h(\mu;\{Q_k\} \cup \P_{X \setminus Q_k}) \geq \h(\mu;\{Q_{k+1}\} \cup \P_{X \setminus Q_{k+1}})
\end{equation}
and
\begin{equation}\label{druga}
\h(\mu;\{Q_k\} \cup \P_{X \setminus Q_k}) \geq \h(\mu;\{Q\} \cup \P_{X \setminus Q}).
\end{equation}

Let $k \in \N$ be arbitrary. Then from (\ref{zawier1}) and (\ref{zawier2}), we get 
\begin{equation}
\h(\mu;\{Q_k\} \cup \P_{X \setminus Q_k}) 
= \sh(\mu(Q_k)) + \sum_{i = 2}^\infty \sh(\mu(P_i \setminus Q_k)) 
\end{equation}
\begin{equation}
= \sh(\mu(Q_k)) + \sum_{i = 2}^k \sh(\mu(P_i \setminus Q)) + \sum_{i=k+1}^\infty \sh(\mu(P_i)) 
\end{equation}
\begin{equation}
= \h(\mu;\{Q_{k+1}\} \cup \P_{X \setminus Q_{k+1}}) + \sh(\mu(Q_k)) - \sh(\mu(Q_{k+1})) 
\end{equation}
\begin{equation}
+ \sh(\mu(P_{k+1})) -\sh(\mu(P_{k+1}\setminus Q)).
\end{equation}
Making use of Observation \ref{shProperty}, we obtain
\begin{equation}
\sh(\mu(Q_k)) + \sh(\mu(P_{k+1})) 
\end{equation}
\begin{equation}
\geq \sh(\mu(Q_{k+1})) + \sh(\mu(P_{k+1}\setminus Q)),
\end{equation}
which proves (\ref{pierwsza}).

To derive (\ref{druga}), we first use inequality (\ref{pierwsza}). Then
\begin{equation}
\h(\mu;\{Q_k\} \cup \P_{X \setminus Q_k}) = \sh(\mu(Q_k)) + \sum_{i = 1}^\infty \sh(\mu(P_i \setminus Q_k)) 
\end{equation}
\begin{equation}
\geq \lim_{n \to \infty} [ \sh(\mu(Q_n)) + \sum_{i = 1}^\infty \sh(\mu(P_i \setminus Q_n)) ]. 
\end{equation}

By (\ref{zawier3}),
\begin{equation}
\lim_{n \to \infty} \sh(\mu(Q_n)) = \sh(\mu(Q)) < \infty.
\end{equation}

To calculate $\lim\limits_{n \to \infty} \sum_{i = 1}^\infty \sh(\mu(P_i \setminus Q_n))$, we will use the Consequence of Lebesgue Theorem. We consider a sequence of functions 
\begin{equation}
f_n: \P \ni P \rightarrow \sh(\mu(P \setminus Q_n)) \in \RR \text{, for } n \in \N.
\end{equation}
 
Let us observe that the Shannon function $\sh$ is increasing on $[0,2^{-\frac{1}{\ln 2}}]$ and decreasing on $(2^{-\frac{1}{\ln 2}},1]$. Thus for a certain $m \in \N$, 
\begin{equation}
\sh(\mu(P_i \setminus Q_n)) \leq 1 \text{, for } i \leq m
\end{equation}
and
\begin{equation}
\sh(\mu(P_i \setminus Q_n)) \leq \sh(\mu(P_i)) \text{, for } i > m,
\end{equation}
for every $n \in \N$. Since $\h(\mu;\P) < \infty$ then
\begin{equation}
\sum_{i = 1}^\infty \sh(\mu(P_i \setminus Q_n)) \leq m + \sum_{i=m+1}^\infty \sh(\mu(P_i)) < \infty.
\end{equation}
Moreover, 
\begin{equation}
\lim_{n \to \infty} \sh(\mu(P \setminus Q_n)) = \sh(\mu(P \setminus Q)),
\end{equation}
for every $P \in \P$.

As the sequence of functions $\{\sh(\mu(P \setminus Q_n))\}_{n \in \N}$ satisfies the assumptions of the Consequence of Lebesgue Theorem then, we get
\begin{equation}
\lim_{n \to \infty} \sum_{i = 1}^\infty \sh(\mu(P_i \setminus Q_n)) = \sum_{i = 1}^\infty \lim_{n \to \infty} \sh(\mu(P_i \setminus Q_n))
\end{equation}
\begin{equation}
= \sum_{i = 1}^\infty \sh(\mu(P_i \setminus Q)) < \infty.
\end{equation}

Consequently, we have
\begin{equation}
\h(\mu;\{Q_k\} \cup \P_{X \setminus Q_k}) \geq \lim_{n \to \infty} [ \sh(\mu(Q_n)) + \sum_{i = 1}^\infty \sh(\mu(P_i \setminus Q_n)) ] 
\end{equation}
\begin{equation}
= \sh(\mu(Q)) + \sum_{i = 1}^\infty \sh(\mu(P_i \setminus Q)) = \h(\mu;\{Q\} \cup \P_{X \setminus Q}),
\end{equation}
which completes the proof.
\end{IEEEproof}

We are ready to summarize the analysis of Algorithm II.1. We present it in the following two theorems.
\begin{theorem}
Let $\Q$ be an error-control family on $X$ and let $\P$ be a $\Q$-acceptable partition of $X$. Family $\R$ constructed by the Algorithm II.1 is a partition of $X$. 
\end{theorem}
\begin{IEEEproof}
Directly from the Algorithm II.1, we get that $\R$ is countable family of pairwise disjoint sets. 

Let us assume that $\R=\{R_i\}_{i=1}^\infty$, since the case when $\R$ is finite family is straightforward. To prove that 
\begin{equation}
\mu(X \setminus \bigcup_{i=1}^\infty R_i) = 0, 
\end{equation}
we will use the Consequence of Lebesgue Theorem.

For every $n \in \N$, we define a function $f_n: \P \rightarrow \RR$ by 
\begin{equation}
f_n(P) := \mu(P \setminus \bigcup_{i=1}^n R_i) \text{, for } P \in \P.
\end{equation}

Clearly, 
\begin{equation}
f_n(P) \leq \mu(P) \text{, for } n \in \N
\end{equation}
and 
\begin{equation}
\sum_{P \in \P} \mu(P) \leq 1.
\end{equation}

To see that the sequence $\{f_n\}_{n = 1}^\infty$ is pointwise convergent, we apply the indirect reasoning. Let $P \in \P$ and let $\e > 0$ be such that, for every $n \in \N$,
\begin{equation}
f_n(P) = \mu(P \setminus \bigcup_{i=1}^n R_i) > \e.
\end{equation}
We put $n:=\lceil\frac{1}{\e}\rceil$. We assume that we have already chosen sets $R_1,\ldots,R_n \in \R$. Since $\mu(P \setminus \bigcup\limits_{i=1}^n R_i) > \e$ then $\mu(R_i) > \e$, for every $i \leq n$. Hence, we have
\begin{equation}
\mu(\bigcup_{i=1}^n R_i) = \sum_{i=1}^n \mu(R_i) \geq n \e  \geq 1,
\end{equation}
as $\R$ is a family of pairwise disjoint sets. Consequently,
\begin{equation}
\mu(P \setminus \bigcup_{i=1}^n R_i) \leq 0,
\end{equation}
which is the contradiction. The sequence $\{f_n\}_{n = 1}^\infty$ is convergent.

Finally, making use of Lebesgue Theorem, we obtain
\begin{equation}
\mu(X \setminus \bigcup_{i=1}^\infty R_i) = \lim_{n \to \infty} \mu(X \setminus \bigcup_{i=1}^n R_i)
\end{equation}
\begin{equation} 
=\lim_{n \to \infty} \sum_{P \in \P} \mu(P \setminus \bigcup_{i=1}^n R_i) = \sum_{P \in \P} \lim_{n \to \infty} \mu(P \setminus \bigcup_{i=1}^n R_i)
\end{equation}
\begin{equation}
= \sum_{P \in \P} \mu(P \setminus \bigcup_{i=1}^\infty R_i) = 0.
\end{equation}
\end{IEEEproof}

\begin{theorem}
Let $\Q$ be an error-control family on $X$ and let $\P$ be a $\Q$-acceptable partition of $X$. Partition $\R$ built by Algorithm II.1 satisfies: 
\begin{equation} \label{ost}
\h(\mu;\R) \leq \h(\mu;\P).
\end{equation}
\end{theorem}
\begin{IEEEproof}
If $\h(\mu;\P) = \infty$ then the inequality (\ref{ost}) is straightforward. Thus let us discuss the case when $\h(\mu;\P) <\infty$.

We denote $\P=\{P_i\}_{i=1}^\infty$, since at most countable number of elements of partition can have positive measure (the case when $\P$ is finite follows similarly). We will use the notation introduced in Algorithm II.1. 

Directly from Lemma \ref{lemma}, we obtain
\begin{equation}
\h(\mu;\P_{X_k}) \geq \h(\mu;\P_{X_{k+1}} \cup \{R_k\}) \text{, for } k \in \N.
\end{equation}
Consequently, for every $k \in \N$, we get
\begin{equation} \label{znowDruga}
\h(\mu;\bigcup_{i = 1}^k \{R_i\} \cup \P_{X_k}) \geq \h(\mu;\bigcup_{i = 1}^{k+1} \{R_i\} \cup \P_{X_{k+1}}).
\end{equation}

Our goal is to show that
\begin{equation}
\h(\mu;\bigcup_{i = 1}^k \{R_i\} \cup \P_{X_k}) \geq \h(\mu;\R),
\end{equation}
for every $k \in \N$.

Making use of (\ref{znowDruga}), we have
\begin{equation}
\h(\mu;\bigcup_{i = 1}^k \{R_i\} \cup \P_{X_k}) 
\end{equation}
\begin{equation}
= \sum_{i=1}^k \sh(\mu(R_i)) + \sum_{i=1}^\infty \sh(\mu(P_i \setminus \bigcup_{j=1}^k R_j))
\end{equation}
\begin{equation}
\geq \lim_{n \to \infty} [ \sum_{i=1}^n \sh(\mu(R_i)) + \sum_{i=1}^\infty \sh(\mu(P_i \setminus \bigcup_{j=1}^n R_j)) ],
\end{equation}
for every $k \in \N$.

We will calculate $\lim\limits_{n \to \infty} \sum\limits_{i=1}^\infty \sh(\mu(P_i \setminus \bigcup\limits_{j=1}^n R_j))$ using the Consequence of Lebesgue Theorem for a sequence of functions $\{f_n\}_{n=1}^\infty$, defined by
\begin{equation}
f_n: \P \ni P \rightarrow \sh(\mu(P \setminus \bigcup_{j=1}^n R_j)) \in \RR \text{, for } n \in \N.
\end{equation} 

Similarly to the proof of Lemma \ref{lemma}, we may assume that there exists $m \in \N$ such that
\begin{equation}
\sh(\mu(P_i \setminus \bigcup_{j = 1}^n R_j)) < 1 \text{, for } i \leq m
\end{equation}
and 
\begin{equation}
\sh(\mu(P_i \setminus \bigcup_{j = 1}^n R_j)) < \sh(\mu(P_i)) \text{, for } i > m,
\end{equation}
for every $n \in \N$. Moreover, 
\begin{equation}
\lim_{n \to \infty} \sh(\mu(P \setminus \bigcup_{j=1}^n R_j)) = \sh(\mu(P \setminus \bigcup_{j=1}^\infty R_j)) = 0,
\end{equation}
for every $P \in \P$ since $\R$ is a partition of $X$. 

Making use of the Consequence of Lebesgue Theorem, we get
\begin{equation}
\lim_{n \to \infty} \sum_{i=1}^\infty \sh(\mu(P_i \setminus \bigcup_{j=1}^n R_j)) = \sum_{i=1}^\infty \sh(\mu(P_i \setminus \bigcup_{j=1}^\infty R_j)) = 0.
\end{equation}

Consequently, for every $k \in \N$, we have
\begin{equation}
\h(\mu;\bigcup_{i = 1}^k \{R_i\} \cup \P_{X_k}) 
\end{equation}
\begin{equation}
\geq \lim_{n \to \infty} [ \sum_{i=1}^n \sh(\mu(R_i)) + \sum_{i=1}^\infty \sh(\mu(P_i \setminus \bigcup_{j=1}^n R_j)) ]
\end{equation}
\begin{equation}
= \sum_{i=1}^\infty \sh(\mu(R_i)) = \h(\mu;\R),
\end{equation}
which completes the proof.
\end{IEEEproof}

\section{Concluding Remarks}

We have seen that in computing the entropy with respect to the error-control family $\Q$ it is sufficient to consider only partitions constructed from the sigma algebra generated by $\Q$. Thus, we may rewritten the definition of the entropy with respect to $\Q$:
\begin{corollary}\label{def}
We have:
\begin{equation}
\begin{array}{rcl}
\H(\mu;\Q) & = & \inf\{  \h(\mu;\P) \in [0;\infty] : \\[0.4ex]
& & \P \text{ is a partition, } \P \prec \Q \text{ and } \P \subset \Sigma_\Q\}.
\end{array}
\end{equation}
\end{corollary}

Let us observe that Algorithm II.1 shows how to find a $\Q$-acceptable partition with the entropy arbitrarily close to $\H(\mu;\Q)$:
\begin{corollary}
Let $\Q$ be an error-control family of $X$. For any number $\e >0$, there exists partition $\P \subset \Sigma_\Q$ such that
\begin{equation}
\h(\mu;\P) \leq \H(\mu;\Q) +\e.
\end{equation} 
\end{corollary}
\begin{IEEEproof}
For simplicity let us assume that $\Q:= \{Q_i\}_{i=1}^\infty$ (the case when $\Q$ is finite or uncountable follows in similar way). Then the partition $\P$, which satisfies the assertion, is of the form:
\begin{equation}
\P := \bigcup_{i=1}^\infty \{ Q_{\sigma(i)} \setminus \bigcup_{k < i} Q_{\sigma(k)} \},
\end{equation}
for specific permutation $\sigma$ of natural numbers. 
\end{IEEEproof}

When $\Q$ is a finite family then the entropy of $\Q$ is always attained on some partition $\P \subset \Sigma_\Q$. More precisely, we have:
\begin{corollary} \label{corAttained}
Let $\Q$ be $n$ element error-control family, where $n \in \N$. Then there exist sets $Q_1,\ldots,Q_k \in \Q$, for specific $k \leq n$, such that
\begin{equation}
\H(\mu;\Q) = \h(\mu;\{Q_i \setminus \bigcup_{j=1}^i Q_j \}_{i=1}^k ).
\end{equation}
\end{corollary}

To see that the entropy with respect to arbitrary, possibly infinite, error-control family does not have to be attained on any partition, we use trivial example from \cite[Example II.1]{Sm}:
\begin{example} \label{exAtt}
Let us consider the open segment $(0,1)$ with sigma algebra generated by all Borel subsets of $(0,1)$, Lebesgue measure $\lambda$ and error control family, defined by 
\begin{equation}
\Q = \{[a,b]: 0 < a < b < 1\}.
\end{equation}
One can verify that $\H(\lambda; \Q) = 0$ but clearly $\h(\lambda;\P) > 0$, for every $\Q$-acceptable partition $\P$.
\end{example}

As an open problem we leave the following question: 
\begin{problem}
Let $\Q$ be an error-control family. We assume that if there exists $\{Q_i\}_{i \in \N} \subset \Q$ such that $Q_k \subset Q_{k+1}$, for every $k \in \N$, then also $\bigcup\limits_{Q \in \Q} Q \in \Q$. We ask if the entropy with respect to $\Q$ is realized by some $\Q$-acceptable partition $\P \subset \Sigma_\Q$.
\end{problem}


\bibliographystyle{IEEEtran}
\bibliography{IEEEfull,entropy}

\end{document}